\documentclass[prd,preprint,a4paper,showpacs,superscriptaddress,nofootinbib,11pt]{revtex4}
 
\usepackage{amssymb}
\usepackage{amsfonts}
\usepackage{amsmath}
\usepackage{hyperref}
\usepackage[dvips]{graphicx}
 
\begin{document}

\date{\today}
 
\title{Testing and selection of cosmological models with $(1+z)^6$ corrections}
 
\author{Marek Szyd{\l}owski}
\email{uoszydlo@cyf-kr.edu.pl}
\affiliation{Department of Theoretical Physics,
Catholic University of Lublin, Al. Rac{\l}awickie 14, 20-950 Lublin, Poland}
\affiliation{Marc Kac Complex Systems Research Centre, Jagiellonian University,
ul. Reymonta 4, 30-059 Krak{\'o}w, Poland}
 
\author{W{\l}odzimierz God{\l}owski}
\email{godlows@oa.uj.edu.pl}
\affiliation{Astronomical Observatory, Jagiellonian University
\\ ul. Orla 171, 30-244 Krak{\'o}w, Poland}
 
\author{Tomasz Stachowiak}
\email{toms@oa.uj.edu.pl}
\affiliation{Astronomical Observatory, Jagiellonian University
\\ ul. Orla 171, 30-244 Krak{\'o}w, Poland}
 
\begin{abstract}
In the paper we check whether the contribution of $(-)(1+z)^6$ type in
the Friedmann equation can be tested. We consider some astronomical
tests to constrain the density parameters in such models.
We describe different interpretations of such an additional term: geometric
effects of Loop Quantum Cosmology, effects of braneworld cosmological models,
non-standard cosmological models in metric-affine gravity, and models with
spinning fluid. Kinematical (or geometrical)
tests based on null geodesics are insufficient to
separate individual matter components when they behave like perfect fluid and
scale in the same way. Still, it is possible to measure their overall
effect. We use recent measurements of the coordinate distances from the
Fanaroff-Riley type IIb (FRIIb) radio galaxy (RG) data, supernovae type Ia
(SNIa) data, baryon oscillation peak and cosmic microwave background radiation
(CMBR) observations to obtain stronger bounds for the contribution of the type
considered. We demonstrate that, while $\rho^2$ corrections are very small,
they can be tested by astronomical observations -- at least in principle.
Bayesian criteria of model selection (the Bayesian factor, AIC, and BIC) are
used to check if additional parameters are detectable in the present epoch.
As it turns out, the $\Lambda$CDM model is favoured over the bouncing model
driven by loop quantum effects. Or, in other words, the bounds obtained from
cosmography are very weak, and from the point of view of the present data
this model is indistinguishable from the $\Lambda$CDM one.
\end{abstract}
 
\pacs{98.80.Jk, 04.20.-q}
 
\maketitle

\section{Introduction}
 
Modern astronomy has led us to the cosmological concordance
model (CCM). The interpretation of observational data suggest that we are living
in almost spatially flat, low density, accelerating universe
filled with dust-like matter \cite{Riess:1998cb,Perlmutter:1998np,deBernardis:2000gy}.
In other words, at first sight there is nothing more to the universe than the
simple $\Lambda$CDM model. In this
model, apart from the observed baryonic matter we have two additional components. One is
in the form of non-relativistic and cold dust matter contributing one third
of the total energy density, while the second one, having negative pressure
violating the strong energy condition, is contributing about two thirds
of the total energy density. This second term is interpreted as non-zero
cosmological constant $\Lambda$. It could be called the minimal model in
that it is the simplest to fit the data so well.
 
Despite that, it is obvious such model is far from reality, and much attention
is paid to possible extensions of physics in that area
\cite{Puetzfeld:2005af,Kamionkowski:2002pc}. Especially the early universe, where general
relativity needs closer association with quantum theories, is open to
modelling. In order to preserve the agreement with late epoch observations, the
modifications need to introduce effects negligible at present, and dominating
for early times, that is to say, decreasing with the scale factor $a$, or
increasing with density $\rho$ -- if one assumes an expanding universe.
The Loop Quantum Gravity offers the possibility of the description of the
discreteness of spacetime at the Planck scale. The classical $\Lambda$CDM
should be emergent from the more fundamental quantum model. Singularities in
classical general relativity should be replaced by the quantum description of
the very beginning of the gravity. For example the Loop Quantum Gravity
cosmological model predicts the bounce instead of the initial singularity.
For full description of the universe evolution both the quantum and classical
regimes should be taken into account.
 
Here, we investigate the observational constraints on a contribution of $\rho^2$
type to the Hubble parameter $H^2(z)$, which could also follow from an
equivalent term of type $(-)(1+z)^6$. Because we have no a priori information
which physical theory is valid in the Universe at the Planck epoch all
possibilities should be taken into account. The viable interpretations of the
presence of this term in $H(a)$ come from the loop quantum, braneworld and
non-Riemannian cosmologies.

\subsection{Universe in Loop Quantum cosmology}
 
One of the important, unsolved problem of the modern cosmology is the problem
of initial conditions for the Universe. Because, at least from our point of
view, the Universe is given in one copy the initial conditions cannot be
taken from ``outside''.
 
Many cosmologists argue that problem of initial conditions for the Universe
can be shifted to the Planckian epoch in which quantum effects were crucial.
In this context, the idea that present expansion of the Universe was preceded
by a contracting phase seems to be very attractive. Geometric effects in Loop
Quantum Cosmology (LQC) predict the presence of $\rho^2$ modification to the
Friedmann equation $H^2(a)$ of negative sign \cite{Singh:2005xg,Ashtekar:2006wn} which
modifies the early evolution leading to a bounce in the generic case.
 
The basic formula $H(z)$ which is used in cosmography assumes very special
form in the case of loop quantum cosmology if effects of curvature are
included. In this case effects of curvature cannot be simply included in a
model by adding a noninteracting curvature fluid with energy
density $\rho_k=-\frac{3k}{a^2}$, $p_k=-\frac{\rho_k}{3}$ because loops quantum
effects manifest themselves in the form of a modification to the classical
Friedmann equation. For example if we consider a closed model the explicit
form is given by \cite{Ashtekar:2006wn,Ashtekar:2006uz,Ashtekar:2006rx,Parisi:2007kv}
\begin{equation}
\label{eq:a0}
H^2=\left(\frac{\rho}{3}+\frac{\Lambda}{3}-\frac{1}{a^2}\right)
\left(1-\frac{\rho}{\rho_{\text{crit}}}-\frac{\Lambda}{\rho_{\text{crit}}}
+\frac{3}{a^2\rho_{\text{crit}}}\right)
\end{equation}
 
where $\rho_{\text{crit}}$ is related with Planck energy density
$\rho_{\text{crit}}=\frac{\sqrt{3}}{16\Pi^2\gamma^2}\rho_{\text{Pl}}$,
$\rho_{\text{Pl}}=\frac{2\pi}{hG^2}$ is Planck density and $\gamma \simeq 0.2375$
is the Immirzi parameter. \cite{Singh:2006im}. With $\rho_{\text{crit}} \simeq 0.82 \rho_{\text{pl}}$
\cite{Xiong:2007aj} and $\rho_{\text{pl}}=5.155 \times 10^{96}\frac{\text{kg}}{\text{m}^3}$
it leads to $\rho_{\text{crit}} \simeq 4.19 \times 10^{93}\frac{\text{g}}{\text{cm}^3}$. Please note that only
the first term in parentheses is corresponding right hand side of the Friedmann equation.
Therefore, loop high-energy modifications of general relativity appear in the
second term in parentheses. The classical general relativity limit can be
obtained if $\rho_{\text{crit}}$ goes to infinity and the second term approaches unity.
 
The effective Friedmann equation assumes the form
\begin{equation}
\label{eq:a1}
H^2=\frac{\rho}{3}\left(1-\frac{\rho}{\rho_{\text{crit}}}\right)+\text{other terms}
\end{equation}
The applications of LQG methods to
cosmology \cite{Bojowald:2001xe,Ashtekar:2003hd,Bojowald:2006da} offer the
possibility of investigating the consequences of discreteness of spacetime
on the quantum level.
One can distinguish a quantum bounce from the classical one, for which
$\rho \ll \rho_{\text{crit}}$ and the quantum corrections are negligible.
 
In the next point we pointed out
similar $\rho^2$ modification of the Friedmann equation as in LQC (for
the comparison of brane world scenarios and LQG cosmology see
\cite{Singh:2006im,Singh:2006sg,Piao:2004hr,Lidsey:2006md,Singh:2003au}).
 
Recently Bojowald pointed out the possibility of testing loops quantum
effects using more precise data in the near future \cite{Bojowald:2007ab}.
He pointed out that while quantum corrections from gravity are commonly
expected to be small, a quantum structure of spacetime can produce potentially
observable effects. Martin Bojowald argued that cosmological models of loops
can be in principle testable but there is no concrete proposal in this context.
We indicate some possibilities of such testing.
However, we must remember that the expansion history
$H(z)$ measures only the kinematic variables \cite{Bludman:2007kg}. We express
some scepticism about the possibility of testing LQC effects by using such a
cosmography. In our opinion, testing the large scale structure (extragalactic
matter distribution) of the Universe would be more valuable.

\subsection{Braneworld cosmological models}
 
The braneworld cosmology is based on the idea that the gravitational field
propagates in bulk spacetime while our observable universe is only
a surface (called brane) embedded in the bigger space.
This idea leads to the presence of an additional term in the Friedmann
equation on the brane. (For review of cosmology with extra dimensions see
\cite{Khoury:2001wf,Rizzo:2004kr,Csaki:2004ay,Lue:2005ya}.)
 
In the present paper we consider a version of higher-dimensional cosmology in
the Randall and Sundrum framework \cite{Randall:1999ee,Randall:1999vf}, which
was already tested to some degree with SN Ia observations
\cite{Godlowski:2004pt,Dabrowski:2002tp}.
 
In this cosmology, the Friedmann equation assumes the following form
\cite{Chung:1999zs,Shtanov:2002mb}
\begin{equation}
\label{eq:b1}
H^2=\frac{\Lambda_4}{3}-\frac{k}{a^2}+\frac{8\pi}{3 M_p^2}\rho+
\epsilon \left(\frac{4\pi}{3 M^3_5}\right)^2\rho^2+\frac{C}{a^4}
\end{equation}
where $\Lambda_4$ is the 4 dimensional cosmological constant, $\epsilon=\pm1$
and $C$ is an integration constant whose magnitude as well as sign can
depend on the initial conditions.
In the present paper we take into account the third term of type $\rho^2$.
This term arises from the imposition of a junction condition for the scale factor
of the brane. This condition has simple interpretation - physical matter
fields are confined to the brane. Both negative and positive $\epsilon$
are possible mathematically because $\epsilon$ corresponds to the metric
signature of the extra dimension \cite{Shtanov:2002mb}.
 
Till now even the sign of $\epsilon$ remains an open question. Sahni and Shtanov
\cite{Sahni:2002vs} discussed some consequences of a special choice of $\epsilon=-1$.
On should note that possibility that the parameter has
negative sign  is crucial because in this case models with timelike extra
dimension can avoid initial singularity by bounce \cite{Brown:2004cs}. The
presence of (negative) $\rho^2$ term in the right hand side of (\ref{eq:b1})
lads to a contracting universe to bounce instead of big-bang curvature
singularity: $\rho \to \infty$, $R_{abcd}R^{abcd} \to \infty$. If the
bounce take place then we have constraint on the value of brane tension
$\sigma$: $|\sigma| \ge 1 \text{MeV}$ because the bounce takes place at densities
greater than during the nucleosynthesis. Note that from observational point of
view the modification of type $(-)\rho^2$ were also recently investigated
by authors in the context of two-brane model \cite{Cline:2002ht} and this
corrections also arises in classically constrained gravity
\cite{Gabadadze:2005ch,McInnes:2005sa}.
 
The fifth term in (\ref{eq:b1}) (called dark radiation) we put
here equal to zero. Note that this term would decay as rapidly as $a^{-4}$
and would  not modify the early evolution as significantly as the third term.
However, please note that non zero dark radiation term can have non-negligible
impact on nucleosynthesis \cite{Ichiki:2002eh,Ichiki:2002yp}.

\subsection{Non-Riemannian cosmological models}
 
The cosmological models basing on the so called non-Riemannian extension of
general relativity theory have been studied for long time (for a review see
\cite{Puetzfeld:2003,Puetzfeld:2004yg,Puetzfeld:2005af}).
Trautman introduced the first non-Riemannian cosmological models
based on Einstein-Cartan theory \cite{Trautman:2006fp}. They are
a modification of general relativity theory by adding the torsion of the
spacetime. In such models, a consequence of spin and torsion is adding
to the Friedmann equation an additional term $\rho^2 \propto a^{-6}$
The main advantage of such a model is
that the problem of initial singularity can be
avoided due to spin effects. Among the different extensions of General
Relativity, the so called metric-affine gravity (MAG) is recently being
studied. In contrast to Riemann-Cartan theory, the connection is not longer
metric which implies that covariant derivative of the metric does not vanish.
 
Recently some authors \cite{Puetzfeld:2004df,Puetzfeld:2004sw, Krawiec:2005jj}
have used the magnitude-redshift relation and observations of distant
supernovae type SNIa to constrain the model parameters. In the case considered
the modified Friedmann equation assumes the following form
\begin{equation}
\label{eq:c1}
H^2=\frac{\rho}{3}-\frac{k}{a^2}-\frac{\Lambda}{3}+v\frac{\psi^2}{a^6},
\end{equation}
where the new constant $v$ can be both negative or positive.
The above equation can be rewritten to a new form:
\begin{equation}
\label{eq:c2}
\Omega_{k,0}+\Omega_{\text{m},0}+\Omega_{\Lambda,0}+\Omega_{\psi,0}=1,
\end{equation}
where $\Omega_{\psi,0} \equiv v \frac{\psi^2}{H^2}$,
$\Omega_{\psi}=\Omega_{\psi,0}a^{-6}$ and $\psi=\mathrm{const}$
is the integration constant.
Therefore non-Riemannian quantities in this model (torsion and non metricity)
will modify very early stages of evolution and are negligible at late times.
The additional density parameter $\Omega_{\psi,0}$ is related to the
non-Riemannian structure.
 
Different astronomical observations from Fanaroff-Riley Type IIb radio
galaxies (FRIIb RG), X ray gas mass fraction to SNIa data can be used to
constrain the present value of the parameter $\Omega_{\psi}$
($\Omega_{\psi}(z=0)=\Omega_{\psi,0}$). This gives limits to the possible
non-Riemannian structure of the spacetime.
 
The above are main possible interpretations of the presence of a $\rho^2$
contribution in the Friedmann equation. We must remember that
cosmography which bases on the behaviour of null geodesics maps the geometry
and kinematics of the Universe in terms of $H(z)$ without reference to the
particular structure of each contribution. Because it measures only average
properties of matter density it is possible to come up with different
forms of contribution leading to the same form $(-)(1+z)^6$ \cite{Bludman:2006cg}.
It was also showed in \cite{Bludman:2006cg} that in the generic case, alpha varying models
lead to a bouncing universe. Interestingly, similar $a^{-6}$ modifications
have also be obtained in the universes with varying constants \cite{Barrow:2004ad}.
 
All the models under consideration can be represented in terms of density
parameters $\Omega_i$. For example for the brane models we can use a set of
parameters
\begin{equation}
\label{eq:2}
\Omega_{k}=-\frac{k}{a^2H^2}=-\frac{k}{\dot{a}^2}, \qquad
\Omega_{\text{m}}=\frac{\rho}{3H^2}, \qquad
\Omega_{\Lambda}=\frac{\Lambda_4}{3H^2}, \qquad
\Omega_{dr}=\frac{a^4}{3H^2}, \qquad
\Omega_{\text{mod}}=-\frac{\rho^2}{3H^2\rho_{cr}}
\end{equation}
where $\sigma$ is the brane tension.
Then equation (\ref{eq:b1}) can be rewritten to the form \cite{Szydlowski:2002sa}
\begin{equation}
\label{eq:3}
\sum \Omega_{i}=1
\end{equation}
Note that $\Omega_{\text{m}}$ and $\Omega_{\text{mod}}$ are not independent, i.e.
$\Omega_{\text{mod},0}=-\frac{\Omega^2_{\text{m},0}}{\Omega_{\text{loops},0}}$,
$\Omega_{\text{loops},0}=-\frac{\rho_{\text{cr}}}{3H_0^2}$. Moreover ${\Omega_{\text{loops},0}}$
is fixed from theory if $H_0$ is known. With $H_0=65$ km s${}^{-1}$ Mpc${}^{-1}$
and $\Omega_{\text{m},0} \simeq 0.3$ it gives
$\Omega_{\text{loops},0} \simeq 5.24 \times 10^{122}$ which gives a contribution
to $\Omega_{\text{mod},0} \simeq  1.72 \times 10^{-124}$ only. It would be
worth mentioning that there are other sources of correction which
act as stiff matter for example self-gravitational corrections of an Ads black
hole \cite{Setare:2006mk}.
 
In our paper we consider standard matter with energy density $\rho$ and
pressure $p=w\rho$, $w=\mathrm{const}$ rather than matter in the form of
non-minimally coupled to gravity scalar field. Bojowald advocated the latter  
approach where matter emerges from the scalar field \cite{Bojowald:2007bg}. 
We assume that standard matter satisfies the
conservation condition which will determine the dependence of energy density
on scale factor (or redshift). This means that we are not considering the
inverse volume effects that would modify the scalar field energy and then
conservation conditions. Good news for our estimation is that for the case
of closed model inverse volume effects are negligible if universe is on
macroscopic level \cite{Parisi:2007kv}.
 
\begin{figure}
\begin{center}
\includegraphics[width=0.9\textwidth]{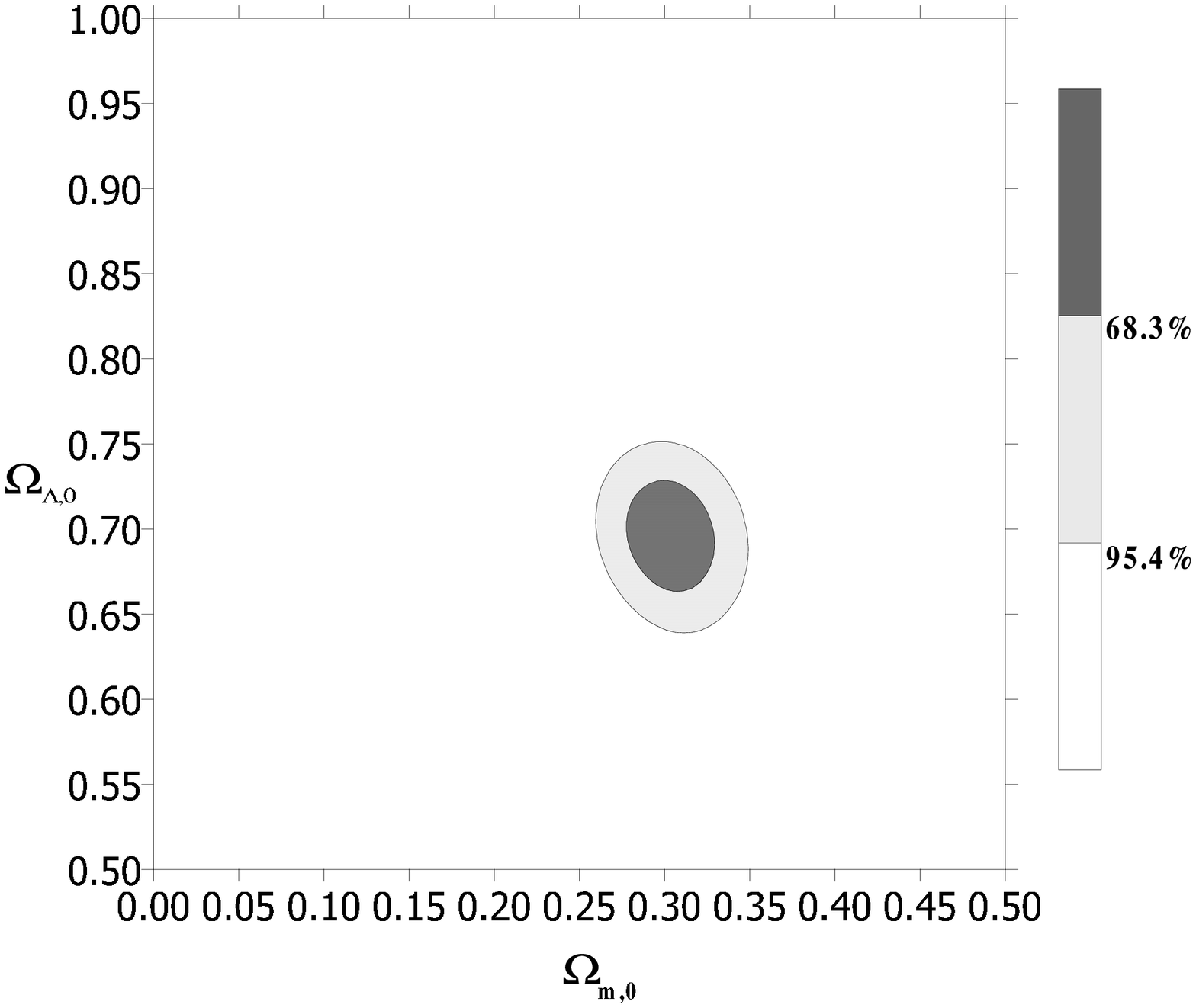}
\end{center}
\caption{The $68.3\%$ and $95.4\%$ confidence levels (obtained from combined
analysis of SN+RG+SDSS+CMBR) on the ($\Omega_{\text{m},0},\Omega_{\Lambda,0}$)
plane.}
\label{fig:3}
\end{figure}

\section{Observational constraints on the FRW bouncing model's parameters}
 
In comparison of the model with observational data we consider two different
strategies. First of all, because the origin of the $\rho^2$ term is not
a priori known we fitted a general $\Omega_{\text{mod},0}$ parameter from the
observations. Accordingly, we obtained that it is too small for detection
(probing) by astronomical observations (cosmography). In the second approach,
we took into account the fact that in the LQG, the density parameter
$\Omega_{\text{mod},0}$ is fixed. We consequently obtained that the model
considered is indistinguishable from the $\Lambda$CDM model. It is a
consequence of the theory itself determining the additional parameter's value
to be very small.
 
Cosmological models are frequently tested against supernovae observations using
the luminosity distance $d_L$ of the Ia supernovae as a function of
redshift \cite{Riess:1998cb}. With these types of tests
for distant SNIa, we can directly observe not the luminosity distance
$d_L$ but their apparent magnitude $m$ and redshift $z$. Taking into account
the fact that absolute magnitude $\mathcal{M}$ of the
supernovae is related to its absolute luminosity $L$, we can obtain
the following relation between the distance modulus $\mu$,
the luminosity distance, the observed magnitude $m$ and the absolute
magnitude $M$
\begin{equation}
\label{eq:11}
\mu \equiv  m - M = 5\log_{10}d_{L} + 25=5\log_{10}D_{L} + \mathcal{M}
\end{equation}
where $D_{L}=H_{0}d_{L}$ and $\mathcal{M} = - 5\log_{10}H_{0} + 25$.
 The luminosity distance of a supernova is a function of
redshift and can be computed from the formulae:
\begin{equation}
\label{eq:12}
d_L(z) =  (1+z) \frac{c}{H_0} \frac{1}{\sqrt{|\Omega_{k,0}|}} \mathcal{F}
\left( H_0 \sqrt{|\Omega_{k,0}|} \int_0^z \frac{d z'}{H(z')} \right)
\end{equation}
where
\begin{equation}
\label{eq:12a}
\left(\frac{H}{H_0}\right)^2= \Omega_{\text{m},0}(1+z)^{3}+\Omega_{k,0}(1+z)^{2}+
\Omega_{\text{mod},0}(1+z)^{6}+\Omega_{\Lambda,0},
\end{equation}
$\Omega_{k,0} = - \frac{k}{H_0^2}$ and
$\mathcal{F} (x) \equiv (\sinh (x), x,\sin (x))$ for $k<0, k=0, k>0$,
respectively.
 
We discuss, for greater generality, non-flat models and keep in mind that for LQG
only the results obtained with fixed, zero $k$ apply. This happens because LQG
models' simple correction of the $\rho^2$ type appear in spatially flat
universes, and assume a more complicated form for $k=\pm 1$ \cite{Bojowald:2001xe,Vandersloot:2006ws}.
 
Daly and Djorgovski \cite{Daly:2003iy} (see also \cite{Zhu:2004ij,Puetzfeld:2004sw,Godlowski:2006vf})
suggest including in the analysis not only supernovae but also radio
galaxies. They pointed out that for tests based on radiogalaxies it is
useful to apply not the luminosity distance $d_L$ but the coordinate distance
$y(z)$ \cite{Weinberg:1972}. The relation between the luminosity distance $d_L$
and the coordinate distance $y(z)$ has the following form
\begin{equation}
\label{eq:12b}
y(z)=\frac{H_0 d_L(z)}{c(1+z)}.
\end{equation}
 
Daly and Djorgovski \cite{Daly:2004gf} have compiled a sample comprising the data
on $y(z)$ for 157 SNIa in the Riess et~al. \cite{Riess:2004nr} Gold dataset
and 20 FRIIb radio galaxies. In our data sets we also include 115 SNIa
compiled by Astier et~al. \cite{Astier:2005qq}.
 
>From the comparison of eq.~(\ref{eq:12a}) and (\ref{eq:12b}) it is easy to
see that the coordinate distance $y(z)$ does not depend on the value of $H_0$.
Unfortunately, we do not observe the coordinate distance $y(z)$ of SNIa
directly. This
distance must be computed from the luminosity distance (or the distance
modulus $\mu$). It is clear that for such a computation a knowledge of the
value of $H_0$ is required. For both supernovae samples we choose the values
of $H_0$ which were used in the original papers.
We used the distance modulus presented in Ref.~\cite{Riess:2004nr,Astier:2005qq} for
the calculation of the coordinate distance. For each sample we choose the
values of $H_0$ appropriate to the data sets. For Riess et al.'s Gold sample
we have $h=0.646$ as the best fitted value and this value is used
for calculation of the coordinate distance for SNIa belonging to this sample.
In turn, the value $h=0.70$ was assumed in the calculations
of the coordinate distance for SNIa belonging to Astier et~al.'s sample,
because the distance moduli $\mu$ presented in Ref.~\cite[Tab.~8]{Astier:2005qq}
were calculated with such an arbitrary value of $h=0.70$.
 
The error of the coordinate distance can be computed as
\begin{equation}
\label{eq:12c}
\sigma^2(y_i)=\left(\frac{10^{\frac{\mu_i}{5}}}{c \left(1+z \right) 10^5} \right)^2
\left(\sigma^2(H_0)+\left(\frac{H_0 \ln{10}}{5}\right)^2\sigma^2(\mu_i)\right)
\end{equation}
where $\sigma_i(\mu_i)$ denotes the statistical error of the distance modulus
determination (note that for Astier et~al.'s sample the intrinsic dispersion
was also included) and $\sigma(H_0)= 0.8$ km/s Mpc denotes the error in $H_0$
measurements.
 
With the use of coordinate distance $y(z)$ as our basic quantity it is easy
to include in our analyse two additional constraints which do not depend on
the value of $H_0$ either. These constraints are obtained from extragalactic
analysis. First, we have the baryon oscillation peaks (BOP) detected in the Sloan
Digital Sky Survey (SDSS) Luminosity Red Galaxies \cite{Eisenstein:2005su}.
They found that value of $A$
\begin{equation}
\label{eq:16}
A \equiv \frac{\sqrt{\Omega_{\text{m},0}}}{E(z_1)^{\frac{1}{3}}}
\left(\frac{1}{z_1\sqrt{|\Omega_{k,0}|}}
\mathcal{F} \left( \sqrt{|\Omega_{k,0}|} \int_0^{z_1} \frac{d z}{E(z)} \right)
\right)^{\frac{2}{3}}
\end{equation}
(where $E(z) \equiv H(z)/H_0$ and $z_1=0.35$) is equal to $A=0.469 \pm 0.017$.
The quoted uncertainty corresponds to one standard deviation, where a
Gaussian probability distribution has been assumed.
 
The second constraint which we include in our analysis is the
so called (CMBR) ``shift parameter''
\begin{equation}
\label{eq:17}
R \equiv \sqrt{\Omega_{\text{m},0}} \, y(z_{lss})=
\sqrt{\frac{\Omega_{\text{m},0}}{|\Omega_{k,0}|}}
\mathcal{F} \left(\sqrt{|\Omega_{k,0}|} \int_0^{z_{\text{lss}}} \frac{d z}{E(z)}\right)
\nonumber
\end{equation}
where $R_0=1.716 \pm 0.062$ \cite{Wang:2004py}.
 
In our combined analysis, we can obtain a best fit model by minimizing the
pseudo-$\chi^2$ merit function \cite{Cardone:2005ut}
\begin{equation}
\label{eq:18}
\chi^{2}=\chi_{\text{SN+RG}}^{2}+\chi_{\text{SDSS}}^{2} +\chi_{\text{CMBR}}^{2}=
\nonumber
\end{equation}
\begin{equation}
\sum_{i}\left(\frac{y_{i}^{\text{obs}}-y_{i}^{\text{th}}}{\sigma_{i}(y_{i})}\right)^{2}+
\left(\frac{A^{\text{mod}}-0.469}{0.017}\right)^{2}+
\left(\frac{R^{\text{mod}}-1.716}{0.062}\right)^{2},
\end{equation}
where $ A^{\text{mod}}$ and $R^{\text{mod}}$ denote the values of $A$ and $R$
obtained for a particular set of the model parameter.
For Astier et al.'s SNIa sample \cite{Astier:2005qq} an additional error in $z$
measurements was taken into account. Here $\sigma_{i}(y_{i})$ denotes the
statistical error (including the error in $z$ measurements) of the coordinate
distance determination.
 
We can obtain constraints for the cosmological parameters by minimizing
the following likelihood function $\mathcal{L}\propto \exp(-\chi^{2}/2)$.
One should note that when we are interested in constraining a particular
model parameter, the likelihood function marginalized over the remaining
parameters of the model should be considered \cite{Cardone:2005ut}.
 
Our results are presented in Table~\ref{tab:1}, Table~\ref{tab:2}
and Fig.~\ref{fig:3}. Table~\ref{tab:1} refers to the minimum $\chi^2$ method,
whereas Table~\ref{tab:2} shows the results from the marginalized likelihood
analysis. In Table~\ref{tab:3}, Table~\ref{tab:4} we repeated our analysis
with the prior $\Omega_{\text{mod},0} \le 0$.
 
>From our combined analysis (SN+RG+SDSS+CMBR), we obtain as the best fit
a flat (or nearly flat universe) with $\Omega_{\text{m},0} \simeq 0.3$, and
$\Omega_{\Lambda,0} \simeq 0.7$. For the $(1+z)^6$ term  we obtain the
stringent bound
$\Omega_{\text{mod},0} \in (-0.26 \times 10^{-9}, 0.31 \times 10^{-4})$
at the $95\%$ confidence level. These results mean that
the positive value of $\Omega_{\text{mod},0}$ is preferred
($\Omega_{\text{mod},0}>0$), however small negative contribution of $(1+z)^6$
type is also available. Our results shows that in the present epoch
contribution of the dark radiation, if it exists, is small and gives only
small corrections to the $\Lambda$CDM model in the low redshift.
 
The above analysis was performed for any spatial curvature. However, please
note that it holds for non-flat models only if the
Friedmann equation does not change in any other way after introducing the
$\rho^2$ term. Unfortunately, LQG models with non-zero curvature were obtained
for different setup of matter components \cite{Bojowald:2001xe,Vandersloot:2006ws}. This means that
for LQG only the flat case consideration apply.
 
The general basic formula for $H(z)$ for universe with $\Omega_{k,0} \ge 0$
rewritten in terms of dimensionless density
parameters $\Omega_i=\frac{\rho_i}{3H^2_0}$ is
\begin{equation}
\label{eq:12d}
\left(\frac{H}{H_0}\right)^2= \left(\Omega_{\text{m},0}(1+z)^{3}
+\Omega_{k,0}(1+z)^{2}+ \Omega_{\Lambda,0}\right)
\left(1-\frac{\Omega_{\text{m},0}(1+z)^3}{\Omega_{\text{loops},0}}
-3\frac{\Omega_{k,0}(1+z)^2}{\Omega_{\text{loops},0}}\right),
\end{equation}
(where $\Omega_{\text{mod},0}=-\frac{\Omega^2_{\text{m},0}}{\Omega_{\text{loops},0}}$).
One should note that in the case considered $\Lambda$ also enters in non-standard
fashion. It means that in this case we do not have the simple reduction to one 
model with different interpretations of $\Omega_{\text{mod}}$, and the same Friedmann 
equation. Then, we are no longer comparing simple $(1+z)^6$ modifications,
but rather different curved models. This case (under the prior $\Omega_{k,0}>0$) 
are analysed and presented separately.
 
When one wants to compare the results obtained for different models, an
interesting question is: how significant is the improvement of the fit due
to the new model? To answer this question one can use the information criteria.
The most popular information criteria used in everyday statistical practise
are the Akaike information criterion (AIC) \cite{Akaike:1974} and the
Bayesian information criterion (BIC) \cite{Schwarz:1978}. These criteria can
be used also for selection of model parameters by providing the preferred
data fit.
 
Usually, incorporating new parameters increases the quality of the fit.
The question is, if it increases it significantly enough. The information
criteria put a threshold which must be exceeded in order to assert an
additional parameter to be important in explanation of a phenomenon. The
discussion of how high this threshold should be caused the appearing of many
different criteria. For us, it suffices to check whether the AIC and
BIC provide sufficient arguments
for incorporation of the new parameters. The power of using information
criteria of model selection was demonstrated by Liddle \cite{Liddle:2004nh}
and Parkinson et al. \cite{Parkinson:2004yx}. Please note that in any case
some future observational data may give arguments in favour of additional
parameters. Also, theoretical considerations could lead to inclusion of new
parameters, even it is not necessary from the point of view of the present
observations.
 
The AIC is defined in the following way \cite{Akaike:1974}
\begin{equation} \label{eq:111}
\text{AIC} = - 2\ln{\mathcal{L}} + 2d,
\end{equation}
where $\mathcal{L}$ is the maximum likelihood and $d$ is a number of free
model parameters. The best model with a parameter set providing the preferred
fit to the data is the one that minimizes the AIC. It is interesting that
the AIC also arises from an approximate minimization of the Kulbak-Leibler
information entropy \cite{Sakamoto:1986ai}.
 
The BIC introduced by Schwarz \cite{Schwarz:1978} is defined as
\begin{equation} \label{eq:112}
\text{BIC} = - 2\ln{\mathcal{L}} + d\ln{N},
\end{equation}
where $N$ is the number of data points used in the fit. Comparing these
criteria, one should note that the AIC tends to favour models with a large number
of parameters unlike the BIC, which penalizes new parameters more strongly.
It is the reason why the BIC provides a more useful approximation to the full
statistical analysis in the case of no priors on the set of model parameters
\cite{Parkinson:2004yx}. This means that while the
AIC is useful in obtaining upper limit to the number of parameters which
should be incorporated to the model, the BIC is more conclusive. One should
note that only the relative value between the BIC of different
models has statistical significance. The difference of $2$ is treated as
a positive evidence (and $6$ as a strong evidence) against the model with the
larger value of the BIC \cite{Jeffreys:1961,Mukherjee:1998wp}.
If we do not find any positive evidence from information criteria the
models are treated as a identical and eventually additional parameters are
treated as not significant. The using of the BIC seems to be especially
suitable whenever the complexity of reference does not increase with the size
of data set. The problem of classification of the cosmological models on the light of
information criteria on the base of the astronomical data was discussed in
our previous papers
\cite{Godlowski:2005tw,Szydlowski:2005xv,Szydlowski:2005bx,Szydlowski:2006gp,Szydlowski:2006ay,Kurek:2007tb}.
 
Our results are presented in Table~\ref{tab:5}. Please note that
the results of statistical analysis for LQG models with non-zero curvature
(under prior $\Omega_{k,0}>0$) are presented as a separate case.
It is clear that in the
light of information criteria, the $(1+z)^6$ term does not increase the
fit significantly. It confirms, what could be expected, that this term, if
it exist, is small in the present epoch.
 
In the Bayesian framework, the quality of models can be compared with help
of evidence \cite{Jeffreys:1961,Mukherjee:2005tr}. We can define the a posteriori
odds for two models -- $M_i$ and $M_j$ -- the so called Bayes factor $B_{ij}$
\cite{Kass:1995}. If a priori we do not favor any model it reduces to the evidence
ratio. Schwarz \cite{Schwarz:1978} showed that for observations coming from
a linear exponential distribution family in the asymptotic approximation
$N \to \infty$ the logarithm of evidence is given by
\begin{equation} \label{eq:113}
\ln{E}=ln{\mathcal{L}} - \frac{d}{2}\ln{N} +O(1).
\end{equation}
It is easy to show that in this case we have a simple relation between
the Bayes factor and the BIC
\begin{equation} \label{eq:114}
2\ln{B_{ij}} = -(\text{BIC}_i - \text{BIC}_j)
\end{equation}
 
If $B_{ij}$ is greater than $3$ it is considered a positive evidence in favor
of $M_i$ model, while $B_{ij}>20$ gives strong, and $B_{ij}>150$ very strong
evidence in favor of model $M_i$ \cite{Szydlowski:2006ay}. We present our results
in Table~\ref{tab:6}. In all cases we obtain positive evidence in favour of
the $\Lambda$CDM model over the bouncing cosmology with the term $\rho^2$. Only in
the case when additional parameter of the theory (responsible for the term $(1+z)^6$)
is fixed the obtained model is indistinguishable from the $\Lambda$CDM model. It
supports the results obtained with help of the AIC and BIC. This result is valid
both with and without the prior of $\Omega_{\text{mod},0} \le 0$. One should note
that because of a non-Gaussian distribution of the a posteriori PDF function
for $\Omega_{\text{mod},0}$ the results obtain with help of Bayes factor (especially
with the  priors $\Omega_{\text{mod},0} \le 0$) should be treated with caution as
only additional support for results obtain with the AIC and BIC.
 
Please also note that if $\Omega_{\text{mod},0}<0$, then we obtain a bouncing
scenario \cite{Molina-Paris:1998tx,Tippett:2004xj,Szydlowski:2005qb} instead of a big bang.
For $\Omega_{\text{m},0}=0.3$, $\Omega_{\text{mod},0}= -0.26 \times 10^{-9}$ and
$h=0.65$ bounces ($H^2=0$) appear for $z \simeq 260$. In this case, the BBN
epoch never occurs and all BBN predictions would be lost.

\section{Location of CMB peaks and BBN in the MAG model}
 
The results obtain in the previous section lead to the conclusion that we
should obtain stronger constrains for model parameter especially for
$\Omega_{\text{mod},0}$. One should note that for doing so,
it is useful to analyse the location of the first peak in
the CMB power spectrum and the predictions of the BBN. Stronger constraints
for model parameters in the MAG model was obtained in Ref.~\cite{Krawiec:2005jj}.
 
The idea of testing models using the location of the first peak in
the CMB power spectrum is based on the fact that the hotter and colder spots
in the CMB can be interpreted as acoustic oscillation in the primeval plasma
during the last scattering. Peaks in the power spectrum correspond to maximum
density of the wave. In the Legendre multipole space these peaks correspond
to the angle subtended by the sound horizon at the last scattering. Further
peaks correspond to higher harmonics of the principal oscillations. The
locations of these peaks are very sensitive to the variations in the model
parameters. Therefore, the position of the first peak
can be used as another way to constrain cosmological models.
 
In the MAG model, assuming $\Omega_{\text{m},0} = 0.3$ and $h=0.72$ for
the standard $\Lambda$CDM universe, the correct positions of the
first peak was obtained in \cite{Krawiec:2005jj} as $\ell_{1} = 220$.
>From the SNIa data analysis, it was found that the Hubble constant
has lower value. Assuming that $H_{0}=65$ km/s Mpc (or $h=0.65$) and
consider the standard $\Lambda$CDM model, with $\Omega_{\text{m},0}=0.3$,
one gets $\ell_{1} = 225$ \cite{Krawiec:2005jj}.
 
Some discrepancy between the observational and theoretical results was found
in this case. Now it was interesting to check whether the presence of the
fictitious fluid $\Omega_{\psi,0}$ changes the locations of the peaks. If we
choose the $H_0=65$ km/s Mpc then agreement with the observation of the
location of the first peak could be obtained for three non-zero values of
the parameter $\Omega_{\psi,0}$. Two positive and one negative values of
this parameter for which the MAG model is admissible are
$3 \times 10^{-11}$, $7 \times 10^{-14}$ and  $-1.4 \times 10^{-10}$.
 
One should note that agreement with prediction of big-bang nucleosynthesis
(BBN) is also crucial for testing of the model. Of course the big-bang
nucleosynthesis is a very well tested area of cosmology and does not allow
for any significant deviation from the standard expansion law apart from
very early times (i.e., before the onset of BBN). The predictions of standard
BBN are in good agreement with observations of the abundance of light elements.
Therefore, all nonstandard terms added to the Friedmann equation should give
only negligible small modifications during the BBN epoch to leave the
nucleosynthesis process unchanged.
 
In our opinion the consistency with BBN is a crucial issue in the models
where the nonstandard term $a^{-6}$ is added in the Friedmann equation.
It is clear that such a term has either accelerated
($\Omega_{\text{mod},0}>0$) or decelerated ($\Omega_{\text{mod},0}<0$) impact
on the Universe expansion. Going backwards in time this term would become
dominant at some redshift. If it had happened before the BBN epoch, the
radiation domination would have never occurred and all the BBN predictions
would be lost.
 
If we assume that the BBN result are preserved in our model, we
obtain another constraint on the amount of $\Omega_{\psi,0}$. Let us assume
that the model modification is negligibly small during the BBN epoch and
the nucleosynthesis process is unchanged. It means that the contribution of
the term $\Omega_{mod}$ cannot dominate over the radiation term
$\Omega_{{\rm r},0} \approx 10^{-4}$ before the beginning of BBN
($z \simeq 10^{8}$)
\begin{equation}
\label{eq:218}
\Omega_{\text{mod},0}(1+z)^{6} < \Omega_{r,0}(1+z)^{4}
\qquad \Longrightarrow \qquad |\Omega_{\psi,0}| < 10^{-20}.
\nonumber
\end{equation}
 
The values of $\Omega_{\text{mod},0}\propto 10^{-3}$ obtained as best fits in
the SNIa data analysis as well as the smallest nonzero value of
$\Omega_{\psi,0}=7\times 10^{-14}$ calculated in the CMB analysis are
unrealistic in the light of the above result. If we take into consideration
the maximum likelihood analysis of SNIa data we have the possibility that
the value of $\Omega_{mod,0}$ is lower than $|10^{-20}|$ in the $2\sigma$
confidence interval.
 
One could argue that bounds from cosmography and CMB are weaker than
those from nucleosynthesis. However, they are model independent. Of course,
nucleosynthesis is a well tested area in cosmology \cite{Carroll:2001bv} 
but it is described rather
in the terms of standard physics without loops corrections. The first step
toward the description of Big-Bang nucleosynthesis are given in the paper by
Bojowald et al. \cite{Bojowald:2007pc}. The authors demonstrated that several 
correction to the equation of state parameter can arise from classical and 
quantum physics, for example loops quantum gravity allows one to compute 
quantum gravity corrections for Maxwell and Dirac field. So we cannot 
a~priori assumed that the corresponding gravity corrections are negligible
during the Big-Bang nucleosynthesis. In the present authors' opinion, the 
significance of bounds obtained from cosmography and CMB is that they are 
independent from the nucleosynthesis bounds and are related to different 
cosmological epochs.

\begin{table}
\caption{Results of the statistical analysis of our model with $(1+z)^6$
like term obtained from $\chi^2$ best fit. The upper section
of the table represents the constraint $\Omega_{k,0}=0$ (flat model).}
\begin{tabular}{@{}p{4.2cm}rrrrr}
\hline  \hline
sample & $\Omega_{k,0}$ & $\Omega_{\text{m},0}$ & $\Omega_{\text{mod},0}$ & $\Omega_{\Lambda,0}$ & $\chi^2$ \\
\hline
SN             &   -   &$ 0.23$ & 0.020            & 0.757 & 295.7     \\
SN+RG          &   -   &$ 0.26$ & 0.013            & 0.727 & 319.5     \\
SN+RG+SDSS     &   -   &$ 0.28$ & 0.009            & 0.711 & 319.6     \\
SN+RG+SDSS+CMBR&   -   &$ 0.30$ &$0.4\times10^{-8}$& 0.700 & 322.4     \\
\hline
SN             &  0.34 &$ 0.00$ & 0.044              & 0.646 & 295.6     \\
SN+RG          &  0.29 &$ 0.05$ & 0.034              & 0.626 & 319.5     \\
SN+RG+SDSS     & -0.03 &$ 0.28$ & 0.011              & 0.739 & 319.5     \\
SN+RG+SDSS+CMBR&  0.04 &$ 0.30$ &$0.279\times10^{-6}$& 0.660 & 321.4     \\
\hline
\end{tabular}
\label{tab:1}
\end{table}

\begin{table}
\caption{Results of the statistical analysis of our model with $(1+z)^6$
term. The values of the model parameters are obtained from
marginalized likelihood analysis. We present maximum likelihood value with
$68.3\%$ confidence ranges. The upper section of the table represents
the constraint $\Omega_{k,0}=0$
(flat model).}
\begin{tabular}{@{}p{4.2cm}cccc}
\hline  \hline
sample & $\Omega_{k,0}$ & $\Omega_{\text{m},0}$ & $\Omega_{\text{mod},0}$ & $\Omega_{\Lambda,0}$ \\
\hline
SN             &   -   & $0.22^{+0.07}_{-0.08}$ & $0.021^{+0.020}_{-0.016}$                 &$0.75^{+0.06}_{-0.05}$ \\
SN+RG          &   -   & $0.27^{+0.05}_{-0.09}$ & $0.013^{+0.018}_{-0.014}$                 &$0.73^{+0.05}_{-0.05}$ \\
SN+RG+SDSS     &   -   & $0.28^{+0.02}_{-0.02}$ & $0.009^{+0.006}_{-0.006}$                 &$0.71^{+0.02}_{-0.02}$ \\
SN+RG+SDSS+CMBR&   -   & $0.30^{+0.02}_{-0.01}$ & $(0.279\times10^{-8})^{0.15\times10^{-5}}_{-3.01\times10^{-9}}$ &$0.70^{+0.01}_{-0.02}$ \\
\hline
SN             &$ 0.10^{+0.19}_{-0.40}$ &$ 0.00^{+0.34}_{-0.00}$ &$0.024^{+0.016}_{-0.027}$                  &$0.77^{+0.14}_{-0.13}$   \\
SN+RG          &$ 0.14^{+0.18}_{-0.39}$ &$ 0.00^{+0.35}_{-0.00}$ &$0.019^{+0.014}_{-0.026}$                  &$0.73^{+0.14}_{-0.13}$   \\
SN+RG+SDSS     &$-0.03^{+0.05}_{-0.10}$ &$ 0.28^{+0.02}_{-0.02}$ &$0.011^{+0.011}_{-0.009}$                  &$0.74^{+0.02}_{-0.02}$   \\
SN+RG+SDSS+CMBR&$ 0.01^{+0.04}_{-0.04}$ &$ 0.30^{+0.02}_{-0.01}$ &$(0.36\times10^{-6})^{+10.64\times 10^{-6}}_{-0.36\times10^{-6}}$ &$0.69^{+0.04}_{-0.03}$   \\
\hline
\end{tabular}
\label{tab:2}
\end{table}

\section{Conclusion}

In the paper we have studied observational constraints on the FRW models with
$\rho^2$ modifications, and obtained stronger limits on the
magnitude of the term $\Omega_{\text{mod}}$ (scaling like $(1+z)^6$).
We pointed out that astronomical observations allow us to
test the total contributions of a fluid scaling like
$(1+z)^6$ but we cannot separate particular terms of a negative sign.
It is a simple consequence of the fact that cosmography measures only the
``average'' density of matter. However, we showed that some
stringent bounds on the value of this total contribution can be given.
 
There are several interpretations of the presence of the $(1+z)^6$ term in
the Friedmann equation: 1) the brane theory; 2) the non-Riemannian theory of gravity;
and 3) the loop quantum cosmology. All models give rise to a $(1+z)^6$ correction
which effects are important in the very early universe and become unimportant
in later evolution. The loop quantum cosmology is more fundamental theory than
the classical theory. We are looking for the quantum cosmology predictions which
are a~priori in agreement with the classical picture of the present Universe
(we showed the quantum loop effect are negligible at the present epoch).
 
Note that if we consider the phantom $\rho^2$
modification, it is also important in the future evolution of the universe
(Big-Rip singularities) \cite{Sami:2006wj}.
 
We used Bayesian methods of model selection to answer the question: which
cosmological model -- with initial singularity or with bounce -- is promoted
by observational data? We have shown that models with a singularity are a
``more economical'' choice for the Universe but bouncing cosmology cannot be
ruled out even on the $1\sigma$ level. However, because the the posteriori
probability function for $\Omega_{\text{mod},0}$ is strongly
non-Gaussian, probability of bounce, i.e. $P(\Omega_{\text{mod},0}<0)$, is less
than $1\%$. Our observational analysis clearly reflects the important role of
independent observational data which enables us to refine the analysis of
model parameters.
 
The analysis of SNIa data as well as both SNIa and FRIIb radio galaxies
(with and without priors coming from baryon oscillation peaks and CMBR
``shift parameter'') shows that the values of $\chi^2$ statistics are
lower for model with $(1+z)^6$ like term, than for the $\Lambda$CDM
model. On the other hand, information criteria show that including such a term does not
increase the quality of the fit significantly; or, alternatively, that the
quality of the available data is not good enough for fitting this new term.
BIC even favours the $\Lambda$CDM model over our model of bounce, although this
preference is weak. These results lie in agreement with the fact that $(1+z)^6$
term is not significant in the present epoch of the Universe. Moreover, if the
additional parameter of the theory $\Omega_{\text{mod},0}$ is fixed, like for
the LQG model, than we obtained model is indistinguishable from the
$\Lambda$CDM model in the Bayesian framework.
 
The combined analysis of SNIa data and FRIIb radio galaxies using  baryon
oscillation peaks and CMBR ``shift parameter'' gives rise to a concordance
universe model which is almost flat with $\Omega_{{\rm m},0} \simeq 0.3$.
>From the above mentioned combined analysis, we obtain the following constraint for the
term which scales like $(1+z)^6$:
$\Omega_{\text{mod},0} \in (-0.26 \times 10^{-9}; 0.31 \times 10^{-4})$
 
We confirm Bojowald's assertion that effects of quantum gravity can by
potentially tested, but the required bounds lie beyond the possibilities
of high precision cosmology. What we obtain from cosmography is very
weak and other tests may by useful.
We have shown that the analysed scenario is compatible with the most recent
low redshift observations of SN Ia, which are independent of physical processes
in the early universe.
We find the other limits on the value of $\Omega_{\psi,0}$ from
measurements of CMB anisotropies and BBN. We obtain the strongest limits in
this case, namely $\Omega_{\psi,0} \le  10^{-20}$ from BBN.
Of course BBN as well as CMB are a very well tested areas of cosmology which
do not allow for significant and substantial changes.
Still, one must remember that, although consistency with
BBN and CMB is a crucial issue,
in such an approach we a~priori assume that brane models (for example) do not change
the physics of the pre-recombination epochs.
 
The analysis was performed for flat and non-flat models with the tacit
assumption that the curvature does not change the physics in any other way than
introducing an appropriate term in the Friedmann equation. Since this is not
the case for LQG, only the flat case considerations give bounds for the
parameter $\Omega_{\text{loops}}$. Although curved LQG models have been introduced
\cite{Bojowald:2001xe,Vandersloot:2006ws}, they have significantly different matter components,
making it impossible to compare them in our scheme.
 
Our general conclusion is that while cosmography can be generally used to
test $\rho^2$ type contribution, such correction turns out to be very small.
Moreover, because we use $H(z)$ function which probes only the average density,
it is not possible to separate effects of LQG from other effects scaling
like $(1+z)^6$, for example effects of the brane. On should note that
in LQG theory $\Omega_{\text{loops},0}$ is fixed and gives an odd
contribution to
$\Omega_{\text{mod},0} \simeq 1.7 \times 10^{-124}$ only, which is far below the
possibility to test by present cosmography. It mean that any
positive evidence for non zero $(1+z)^6$ term
eventually obtained from cosmography can not be connected with LQG.

\begin{table}
\caption{Results of the statistical analysis of the bouncing model with
the negative $(1+z)^6$ term obtained from $\chi^2$ best fit.
The upper section of the table represents the constraint $\Omega_{k,0}=0$ (flat model).}
\begin{tabular}{@{}p{4.2cm}rrrrr}
\hline  \hline
Sample & $\Omega_{k,0}$ & $\Omega_{\text{\text{m}},0}$ & $\Omega_{\text{mod},0}$ & $\Omega_{\Lambda,0}$ & $\chi^2$ \\
\hline
SN             &   -   &$ 0.31$ & 0.000 & 0.69 & 297.5     \\
SN+RG          &   -   &$ 0.32$ & 0.000 & 0.68 & 320.4     \\
SN+RG+SDSS     &   -   &$ 0.30$ & 0.000 & 0.70 & 322.4     \\
SN+RG+SDSS+CMBR&   -   &$ 0.30$ & 0.000 & 0.70 & 322.5     \\
\hline
SN             & -0.28 &$ 0.43$ & 0.000 & 0.85 & 296.0     \\
SN+RG          & -0.18 &$ 0.39$ & 0.000 & 0.80 & 319.5     \\
SN+RG+SDSS     &  0.06 &$ 0.29$ & 0.000 & 0.65 & 321.1     \\
SN+RG+SDSS+CMBR&  0.00 &$ 0.30$ & 0.000 & 0.70 & 322.5     \\
\hline
\end{tabular}
\label{tab:3}
\end{table}

\begin{table}
\caption{Results of the statistical analysis of the bouncing model with the
negative $(1+z)^6$ term. The values of the model parameters are obtained from
the marginalized likelihood analysis. We present the maximum likelihood values
with $68.3\%$ confidence ranges. The upper section of the table represents
the constraint $\Omega_{k,0}=0$ (flat model).}
\begin{tabular}{@{}p{4.2cm}cccc}
\hline  \hline
Sample & $\Omega_{k,0}$ & $\Omega_{\text{m},0}$ & $\Omega_{\text{mod},0}$ & $\Omega_{\Lambda,0}$ \\
\hline
SN             &   -   & $0.33^{+0.02}_{-0.02}$ & $0.000_{-0.006}$     &$0.67^{+0.02}_{-0.02}$ \\
SN+RG          &   -   & $0.33^{+0.03}_{-0.02}$ & $0.000_{-0.006}$     &$0.67^{+0.02}_{-0.03}$ \\
SN+RG+SDSS     &   -   & $0.31^{+0.02}_{-0.02}$ & $0.000_{-0.002}$     &$0.69^{+0.02}_{-0.02}$ \\
SN+RG+SDSS+CMBR&   -   & $0.30^{+0.02}_{-0.01}$ & $0.00000_{-0.22\times10^{-9}}$ &$0.70^{+0.01}_{-0.02}$ \\
\hline
SN             &$-0.42^{+0.24}_{-0.22}$ &$ 0.52^{+0.09}_{-0.09}$ &$0.000_{-0.011}$    &$0.91^{+0.11}_{-0.15}$   \\
SN+RG          &$-0.36^{+0.25}_{-0.20}$ &$ 0.49^{+0.10}_{-0.11}$ &$0.000_{-0.011}$    &$0.85^{+0.12}_{-0.13}$   \\
SN+RG+SDSS     &$ 0.08^{+0.06}_{-0.05}$ &$ 0.29^{+0.02}_{-0.02}$ &$0.000_{-0.004}$    &$0.63^{+0.05}_{-0.06}$   \\
SN+RG+SDSS+CMBR&$ 0.00^{+0.02}_{-0.03}$ &$ 0.30^{+0.02}_{-0.01}$ &$0.00000_{-0.21\times10^{-9}}$&$0.70^{+0.02}_{-0.02}$   \\
\hline
\end{tabular}
\label{tab:4}
\end{table}

\begin{table}
\caption{The values of AIC and BIC for the
$\Lambda$CDM model and Bouncing Cosmology
model (with the term $(1+z)^6$) without and with priors $\Omega_{\text{mod},0} \le 0$.
Separately the LQG model with fixed $\Omega_{\text{mod},0}$ ($\Omega_{k,0} > 0$) is considered.
The upper section of the table represents the constraint $\Omega_{k,0}=0$
(flat model).}
\begin{tabular}{c|cc|cc|cc|cc}
\hline \hline
\multicolumn{1}{c}{}&
\multicolumn{2}{c}{$\Lambda$CDM}&
\multicolumn{2}{c}{$\Lambda$BCDM}&
\multicolumn{2}{c}{$\Lambda$BCDM($\Omega_{\psi,0} \le 0$)}&
\multicolumn{2}{c}{LQG}\\
\hline \hline
sample & AIC & BIC & AIC & BIC & AIC & BIC & AIC & BIC \\
\hline
SN             & 299.5& 303.1& 299.7 & 306.9 & 301.5 & 308.7 & 299.5& 303.1 \\
SN+RG          & 322.4& 326.1& 323.5 & 330.9 & 324.4 & 331.8 & 322.4& 326.1 \\
SN+RG+SDSS     & 324.4& 328.1& 323.6 & 332.0 & 326.4 & 333.8 & 324.4& 328.1 \\
SN+RG+SDSS+CMBR& 324.5& 328.2& 326.4 & 333.8 & 326.5 & 333.9 & 324.5& 328.2 \\
\hline
SN             & 300.0& 307.2& 301.6 & 312.4 & 302.0 & 312.8 & 301.5& 308.7 \\
SN+RG          & 323.5& 330.9& 325.5 & 336.5 & 325.5 & 336.5 & 324.4& 331.8 \\
SN+RG+SDSS     & 325.1& 332.5& 325.5 & 336.5 & 327.1 & 338.1 & 325.1& 332.5 \\
SN+RG+SDSS+CMBR& 326.5& 333.9& 327.4 & 338.4 & 328.5 & 339.5 & 326.5& 333.9 \\
\hline
\end{tabular}
\label{tab:5}
\end{table}
 
\begin{table}
\caption{The values of Bayes factor for models:
1) $\Lambda$CDM model, 2) Bouncing Cosmology model
(with the term $(1+z)^6$) and 3) Bouncing Cosmology model (with $(1+z)^6$ term)
with priors $\Omega_{\text{mod},0} \le 0$, 4) LQG model with fixed $\Omega_{\text{mod},0}$
($\Omega_{k,0} > 0$).
The upper section of the table represents the constraint $\Omega_{k,0}=0$
(flat model).}
\begin{tabular}{ccccccc}
\hline \hline
\hline \hline
sample & $B_{12}$ & $B_{13}$ & $B_{14}$ & $B_{23}$ & $B_{42}$ & $B_{43}$\\
\hline
SN             &    6.69  &  16.44  & 1.00 &  2.46 &  6.69 & 16.44   \\
SN+RG          &   11.02  &  17.29  & 1.00 &  1.57 & 11.02 & 17.29   \\
SN+RG+SDSS     &    7.03  &  17.29  & 1.00 &  2.46 &  7.03 & 17.29   \\
SN+RG+SDSS+CMBR&   16.44  &  17.29  & 1.00 &  1.05 & 16.44 & 17.29   \\
\hline
SN             &   13.46  &  16.44  & 2.12 &  1.22 &  6.36 &  7.77   \\
SN+RG          &   16.44  &  17.29  & 1.57 &  1.00 & 10.49 & 11.02   \\
SN+RG+SDSS     &    7.39  &  17.29  & 1.00 &  2.23 &  7.39 & 17.29   \\
SN+RG+SDSS+CMBR&    9.49  &  17.29  & 1.00 &  1.73 &  9.49 & 17.29   \\
\hline
\end{tabular}
\label{tab:6}
\end{table}
 
\section*{Acknowledgements}
The work was supported in part by project ``COCOS'' No. MTKD-CT-2004-517186.
The authors also thank Dr. A.G. Riess, Dr. P. Astier and Dr. R. Daly for
the detailed explanation of their data samples.

\section*{Appendix. Basics of Loop Quantum Cosmology.}
 
In this appendix we give some selected information about Loop Quantum Gravity
(LQG) connected to the subject of this work. Our main challenge is to show
how to obtain equation (1) considered in this paper (see also a nice summary
\cite{Magueijo:2007wf}). We concentrate rather on the main steps and for
detailed calculations we direct to the references.
 
Loop Quantum Gravity describe the gravitational field as a $SU(2)$ non-Abelian
gauge field using background independent methods. The canonical fields are so
called Ashtekar variables $(A,E)$ \cite{Ashtekar:1987gu} which take value in
$\mathfrak{su}(2)$ and $\mathfrak{su}(2)^*$ algebras respectively.
These variables are analogues of the four potential and electric field
in electrodynamics. The Ashtekar variables are strictly connected
with triad representation. However in LQG gauge fields describe only
spatial part $\Sigma$ when time is treated separately.
To quantise this theory in the background independent way
ones introduce holonomies of connection $A$
\begin{equation}
h_{\alpha}[A] = \mathcal{P} \exp \int_{\alpha} A  \ \ \text{where 1-form} \ \   A=\tau_i A^i_a dx^a
\label{holo}
\end{equation}
and conjugated fluxes
\begin{equation}
F_{S}^i[E] = \int_S dF^i   \ \ \text{where 2-form} \ \  dF_i =\epsilon_{abc} E^a_i dx^b \wedge dx^c.
\end{equation}
which are background independent observables.
In equation (\ref{holo}) we have introduced $\tau_i =-\frac{i}{2} \sigma_i$
where $\sigma_i$ are Pauli matrices. The quantisation of theory lead to
important result that volumes and areas have discrete spectrum. For further
calculations it will be important to notice that there exists a minimal area
$\Delta= 2\sqrt{3}\pi \gamma l_{\text{Pl}}^2$ \cite{Ashtekar:1996eg}.
For introductory review on loop quantisation see \cite{Nicolai:2005mc}.
 
Dynamics of theory is contained in the scalar constraint
\begin{eqnarray}
H_{\rm G} =  \frac{1}{16 \pi G} \int_{\Sigma} d^3 x N(x) \frac{E^a_iE^b_j}{\sqrt{|\mathrm{det} E|}}
  \left[  {\varepsilon^{ij}}_k F_{ab}^k   -  2(1+\gamma^2)  K^i_{[a}  K^j_{b]} \right]
\label{scalar}
\end{eqnarray}
where $F$ is a field strength $F=dA+\frac{1}{2}[A,A]$ and $K$ is an extrinsic
curvature. Constant $\gamma$ in equation (\ref{scalar}) is so called
Barbero-Immirzi parameter, $\gamma=\ln 2 / (\pi \sqrt{3} )$.
 
We want to show now how to apply LQG to case of the flat FRW model considered in
this paper. The results presented below base on papers
\cite{Ashtekar:2006rx,Ashtekar:2006uz,Ashtekar:2006wn}
where reader can find detailed calculations and analysis.
The FRW $k=0$ spacetime metric can be written as
\begin{equation}
ds^2=-N^2(x) dt^2 + q_{ab}dx^adx^b
\end{equation}
where $N(x)$ is the lapse function (here we choose gauge $N(x)=1$) and the
spatial part of the metric is expressed as
\begin{equation}
q_{ab}= \delta_{ij} {\omega^i_a} {\omega^i_a}= a^2(t) {^oq}_{ab}\delta_{ij}  {^o\omega^i_a}{^o\omega^i_a}.
\end{equation}
In this expression ${^oq}_{ab}$ is fiducial metric and ${^o\omega^i_a}$ are
co-triads dual to the triads ${^oe^a_i}$,  ${^o\omega^i}({^oe_j})=\delta^i_j$
where $^o\omega^i={^o\omega^i_a}dx^a$ and $^oe_i={^oe_i^a}\partial_a$.
In the case considered the Ashtekar variables are
\begin{eqnarray}
A &\equiv& \Gamma +\gamma K  = {c} V_0^{-1/3} \ {^o\omega^i_a}  \tau_i dx^a , \label{A} \\
E &\equiv& \sqrt{|\det q|} e  = {p} V_0^{-2/3} \sqrt{^oq} \ {^oe^a_i} \tau_i  \partial_a \label{E}
\end{eqnarray}
where $V_0$ is volume of fiducial cell and $\Gamma$ is spin connection.
The parameters $(c,p)$ can be since now considered as canonical variables with
fundamental Poisson bracket $\{c,p \} = {8 \pi G \gamma }/{3} $. In quantum
theory $p$ is replaced by operator $\hat{p}$ which acting on the eigenvector
$|\mu \rangle$ gives
\begin{equation}
\hat{p}|\mu \rangle =\mu \frac{8\pi l^2_{\text{Pl}}\gamma }{6} |\mu \rangle .
\label{op1}
\end{equation}
where $\mu\in \mathbb{R}$ and eigenvectors fulfils relation of orthogonality
$\langle \mu_i |\mu_j \rangle = \delta_{\mu_i,\mu_j}$.
There is no well defined operator of variable $c$, instead of this there exists
another fundamental operator defined as
\begin{equation}
\widehat{\exp \frac{i\lambda c}{2}} |\mu \rangle = |\mu +\lambda \rangle.
\label{op2}
\end{equation}
 
With use of definition (\ref{holo}) we can calculate holonomy for connection
(\ref{A}) in particular direction $^oe^a_i\partial_a$
\begin{eqnarray}
h_{i}^{(\lambda)} = \mathbb{I}\cos \left( \frac{\lambda c}{2}\right)+
\tau_i\sin \left( \frac{\lambda c}{2}\right).
\label{hol2}
\end{eqnarray}
Holonomy is well defined operator which acting on vector $|\mu \rangle$ gives
\begin{equation}
\hat{h}_{i}^{(\lambda)}|\mu \rangle  = \frac{1}{2} \left( |\mu +\lambda \rangle + |\mu -\lambda \rangle \right)
                                     + \frac{1}{i} \left( |\mu +\lambda \rangle - |\mu -\lambda \rangle \right) \tau_i
\end{equation}
where we have used definition (\ref{op2}).
>From particular holonomies (\ref{hol2}) we can
construct holonomy along the closed curve $\alpha=\Box_{ij}$.
This holonomy can be written as
\begin{eqnarray}
h_{\Box_{ij}}^{(\mu)} = h_{i}^{(\mu)} h_{j}^{(\mu)} h_{i}^{(\mu)-1} h_{j}^{(\mu)-1}
                      = \mathbb{I} + \mu^2 V_0^{2/3} F^k_{ab} \tau_k {^oe^a_i} {^oe^b_j} +   \mathcal{O}(\mu^3)
\label{deriv}
\end{eqnarray}
and inverting this equation we obtain expression for field strength
\begin{equation}
F^k_{ab} \approx - 2
 \frac{\text{tr}\left[\tau_k \left( h^{(\mu)}_{\Box_{ij}}-\mathbb{I} \right)  \right]}{ {\mu}^2 V_0^{2/3}} {^o\omega^i_a}{^o\omega^j_b}=
  \frac{\sin^2\left({\mu} c \right) }{{\mu}^2 V_0^{2/3}}   \epsilon_{kij} {^o\omega^i_a}{^o\omega^j_b}
\label{lim}
\end{equation}
In the limit ${\mu}\rightarrow 0$ we retrieve classical formula. However this
limit does not exist in quantum theory because of area gap $\Delta$. LQG tell
us that we should stop shrinking the loop $\mu \rightarrow  \bar{\mu}$
when physical area $p \bar{\mu}^2=\Delta$.
Then inserting expressions for field strength (\ref{lim}) and for $A$ and $K$
to (\ref{scalar}) we obtain the effective Hamiltonian with holonomy corrections
\begin{equation}
H_{\text{eff}} =  - \frac{3}{8 \pi G \gamma^2} \sqrt{p} \left[ \frac{ \sin \left( \bar{\mu} c
\right) }{\bar{\mu}} \right]^2 +  p^{3/2}\rho_{\text{m}}
\label{model}
\end{equation}
where we have also added matter part. Additional feature of theory is
Hamiltonian constraint $H_{\text{eff}} =0$ which implies
\begin{equation}
\frac{1}{\gamma^2 p}  \left[ \frac{ \sin \left( \bar{\mu} c \right)  }{\bar{\mu}}   \right]^2 =
\frac{8 \pi G }{3}  \rho_{\text{m}}.
\label{constraint}
\end{equation}
Now we can use the Hamilton equation to determinate dynamics of the canonical
variable $p$
\begin{equation}
\dot{p} = \{p,H_{\text{eff}}  \} = \frac{2}{\gamma} \frac{ \sqrt{p} }{\bar{\mu} }  \sin \left( \bar{\mu} c \right) \cos \left( \bar{\mu} c  \right).
\label{eqatp}
\end{equation}
Combining equations (\ref{constraint}) and (\ref{eqatp}) we finally obtain the
modified Friedmann equation with holonomy correction
\begin{equation}
H^2 = \frac{8\pi G}{3} \rho_{\text{m}} \left(1-\frac{\rho_{\text{m}}}{\rho_{\text{c}}}  \right)
\end{equation}
where we have introduced critical energy density
\begin{equation}
\rho_{\text{c}} = \frac{3}{8\pi G \gamma^2 \bar{\mu}^2 p} =  \frac{3}{8\pi G \gamma^2 \Delta} =
  \frac{\sqrt{3}}{16\pi^2 \gamma^3 l_{\text{Pl}}^4 }.
\end{equation}
 
For further details of Loop Quantum Cosmology we recommend the introductory
review \cite{Bojowald:2006da}.

\end{document}